\documentclass[aps,twocolumn,showpacs,preprintnumbers,amsmath,amssymb,nofootinbib,superscriptaddress]{revtex4}

\usepackage{graphicx}
\usepackage{dcolumn}
\usepackage{bm}
\usepackage{amsmath}
\usepackage{hyperref}

\def\mk{\mathbf{k}}
\def\mq{\mathbf{q}}

\def\sgn{\mathrm{sgn}}
\def\Re{\mathrm{Re}}
\def\Im{\mathrm{Im}}

\begin{document}

\title{Out-of-plane spin polarization of edge currents in Chern insulator with Rashba spin-orbit interaction}

\author{Tsung-Wei Chen}
\email{twchen@mail.nsysu.edu.tw}\affiliation{Department of
Physics, National Sun Yat-sen University, Kaohsiung 80424, Taiwan}

\author{Chin-Lun Hsiao}\affiliation{Department of
Physics, National Sun Yat-sen University, Kaohsiung 80424, Taiwan}
\affiliation{Department of Physics and Center for Theoretical
Sciences, National Taiwan University, Taipei 106, Taiwan}

\author{Chong-Der Hu}
\affiliation{Department of Physics and Center for Theoretical
Sciences, National Taiwan University, Taipei 106, Taiwan}


\date{\today}

\begin{abstract}
We investigate the change in the non-zero Chern number and
out-of-plane spin polarization of the edge currents in a honeycomb
lattice with the Haldane-Rashba interaction. This interaction
breaks the time-reversal symmetry due to the Haldane phase caused
by a current loop at site-I and -II atoms and also accounts for
the Rashba-type spin-orbit interaction. The Rashba spin-orbit
interaction increases the number of Dirac points and the
band-touching phenomenon can be generated by tuning the on-site
potential in the non-zero Haldane phase. By using the Pontryagin
winding number and numerical Berry curvature methods, we find that
the Chern number pattern is $\{+2, -1, 0\}$ and $\{-2, +1, 0\}$
for the positive and negative Haldane phase, respectively. A
non-zero Chern number is called a Chern-insulating phase. We
discovered that changes in both the Haldane phase and on-site
potential leads to a change in the orientation of the bulk spin
polarization of site-I and site-II atoms. Interestingly, in a
ribbon with a zigzag edge, which naturally has site-I atoms at one
outer edge and site-II atoms at the opposite outer edge, the spin
polarization of the edge states approximately obeys the properties
of bulk spin polarization regardless of the change in the Chern
number. In addition, even when the Chern number changes from $+2$
to $-1$ (or $-2$ to $+1$), by tuning the strength of the on-site
potential, the sign of the spin polarization of the edge states
persists.  This approximate bulk-edge correspondence of the spin
polarization in the Haldane-Rashba system would play an important
role in spintronics, because it enables us to control the
orientation of the spin polarization in a single Chern-insulating
phase.


\end{abstract}
 \maketitle

\section{Introduction}\label{sec:Intro}

The classification of matter by using various topological
properties of the band structure has become a central subject
in condensed matter physics. The Berry curvature\cite{Berry1984}
embedded within the band structure plays an important role in
constructing various topological indices.

Thus, it was found that the quantized Hall conductance in the
integer quantum Hall regime can be determined by the topological
index Chern number (Ch) or TKNN integer~\cite{Thouless1982}, which
was shown to be the integral of the Berry curvature over the
Brillouin zone of the bulk system. A change in the Chern number
corresponds to a change in the number of edge currents, which is
the bulk-edge correspondence~\cite{Hatsugai1993}. Importantly, in
Ref.~\cite{Haldane1988}, Haldane proposed a
time-reversal-symmetry-broken two-band model that exhibits an
energy gap without Landau levels in the bulk and gapless
excitation at the edge of a honeycomb lattice, the so-called Chern
insulator. The variation in the Chern number is due to a
band-touching phenomenon, in which the energy gap of the bulk
system closes and opens at some Dirac points in the Brillouin
zone. In this sense, the band-touching phenomenon, generated by
different mechanisms, would lead to different topological phases.
Accounting for the spin-orbit interaction, the combination of a
Chern insulator and its time-reversal pair results in the
quantum-spin Hall phase in graphene (Kane-Mele model)
\cite{Kane2005} and HgTe quantum wells (BHZ model)
\cite{Bernevig2006} in a two-dimensional system. This
time-reversal-invariant quantum spin Hall insulator composed of
two time-reversal-breaking Chern insulators has been generalized
to three-dimensional systems, where it is called a topological
insulator \cite{Fu2007}. This was later confirmed experimentally
\cite{Hsieh2008}. The topological phases of the time-reversal
invariant system can be determined by calculating the $Z_2$
topological index of the bulk system \cite{Hasan2010} or the spin
Chern number \cite{Sheng2006, Prodan2009, Shan2010, Chu2008}. On
the other hand, a quantum anomalous Hall phase can be exhibited in
graphene with a Rashba spin-orbit interaction in the presence of
an exchange field \cite{Qiao2010}.

It was theoretically proposed that in a honeycomb lattice with a
Kane-Mele Hamiltonian including the Rashba spin-orbit interaction,
the quantum spin Hall phase persists in the presence of an
exchange field, i.e., the time-reversal symmetry is broken
\cite{Yang2011, Chen2011}. In Ref.~\cite{Ma2015}, it was
demonstrated that the edge state in the quantum spin Hall phase
can persist even for a strong magnetic field.

Moreover, because the spin-orbit interaction relates to the
orbital motion of the electron, a change in the Chern number from
positive to negative integer leads to a change in the direction of
the edge current, and the out-of-plane spin polarization
(hereafter called spin polarization for convenience) may change
due to the spin-orbit coupling.  However, this is not always true
because the spin polarization of the edge states would also be
affected by the on-site potential, as in the honeycomb lattice.
The edge current is exhibited by the Haldane current loops at
site-I or site-II atoms. The zigzag ribbon naturally has site-I
atoms at one outer edge and site-II at the opposite outer edge.
This implies that the two Haldane current loops must contribute
unequally to the edge currents [see Fig. \ref{figLattice}(a)], and
that one of the two Haldane current loops would dominate at one
edge. At right (left) side of the ribbon, the site-I (site-II)
atoms has more nearest neighbor atoms than site-II (site-I) atoms.
Because the Rashba spin-orbit interaction governs the nearest
neighbor hopping of different sites, this implies that the two
different sites must contribute unequally to the spin polarization
of edge currents. Furthermore, different spin states at the edges
of a honeycomb lattice with a zigzag edge would occupy different
sites under broken time reversal symmetry. Thus, the on-site
potential would behave like an effective magnetic field for the
edge states. As a result, the bulk spin polarization for site-I
and site-II atoms would have the same behavior as the spin
polarization of edge currents. In a manner similar to the
bulk-edge correspondence in the Chern number, there would be an
approximate bulk-edge correspondence for spin polarization.
Therefore, if the band-touching phenomenon is caused by the
on-site potential, a question arises naturally: Can the spin
polarization of the edge currents also be protected in the
time-reversal-symmetry-broken system with a spin-orbit interaction
under a change in the Chern number?

Furthermore, in Ref.~\cite{Shao2008}, the authors proposed a
scheme to realize the Haldane model by using ultracold atoms
trapped in an optical lattice. Later, the detection of the
topological phase of a Chern insulator was proposed in
Ref.~\cite{Liu2013}. Recently, it was shown experimentally that
the Haldane model can be exhibited with ultracold fermions
\cite{Jo2014}.

Motivated by these issues, in this paper, we study a Haldane
system with Rashba spin-orbit interaction \cite{Zarea2009} in a
honeycomb lattice, from now on called a Haldane-Rashba system. The
change in the Chern number in the Haldane system generated by the
on-site potential is known to be $\{-1,0\}$ and $\{+1,0\}$ for a
negative and positive Haldane phase, respectively. A non-zero
Chern number is called a Chern-insulating phase. The number of
Dirac points of the Haldane system is 2 in the Brillouin zone (see
Fig.~\ref{figDps}). On the other hand, the number of Dirac points
in the pure Rashba system is 6 in the honeycomb lattice
\cite{Zarea2009} [see Fig.~\ref{figLattice}(b)]. We will show that
in the Haldane-Rashba system the resulting Chern number patterns
are $\{-2, +1, 0\}$ and $\{+2,-1,0\}$ for a negative and positive
Haldane phase, respectively. Therefore, the Haldane-Rashba system
would be a candidate for the investigation of the persistence of
spin polarization of the edge states under a change in the Chern
number. We will show that the spin-polarization of the edge
currents persists under the change in Chern number and furthermore
that the magnitude of the spin polarization can be tuned by the
on-site potential. This implies that we could control the
orientation of the spin polarization of the edge currents in a
single Chern-insulating phase.

This paper is organized as follows. In section \ref{sec:HRS}, we
introduce the Haldane-Rashba system. We map the honeycomb lattice
to a square lattice and find the Dirac points. In section
\ref{sec:Ch pattern}, we obtain the Chern number pattern of the
Haldane-Rashba system by using the Pontryagin winding number method
and numerically calculating the Berry curvature. In section
\ref{sec:SP}, we study the spin polarization of the bulk and edge
states. The conclusion is given in section \ref{sec:Conclusion}.

\section{Dirac points in the Haldane-Rashba system}\label{sec:HRS}
\begin{figure}
\begin{center}
\includegraphics[width=6cm,height=4cm]{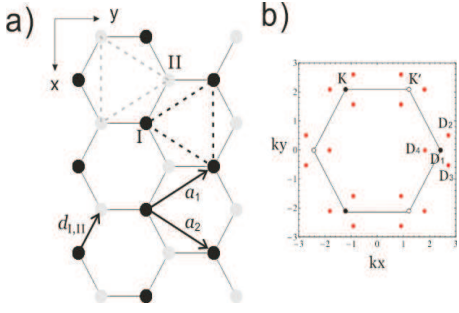}
\end{center}
\caption{(a) Honeycomb lattice and current loop (dashed lines) in
the Haldane model. (b) Brillouin zone and the distribution of the
Dirac points for $\alpha=0.7t$.}\label{figLattice}
\end{figure}

In this section, we study the number of Dirac points in the
Haldane-Rashba system in the square Brillouin zone transformed
from the honeycomb lattice. The Haldane-Rashba Hamiltonian is
given by
\begin{equation}\label{sec:HRS-HRH}
\begin{split}
H=&t\sum_{<i,j>}c_{i\sigma}^{\dag}c_{j\sigma}+t_2\sum_{\ll
i,j\gg,\sigma}e^{-i\nu_{ij}\phi}c_{i\sigma}^{\dag}c_{
j\sigma}\\
&+i\alpha\sum_{<i,j>}\sum_{\sigma\sigma'}(\mathbf{s}\times\mathbf{d}_{ij})^{\sigma\sigma'}_zc^{\dag}_{i\sigma}c_{j\sigma'}+G\sum_{i\sigma}\xi_ic^{\dag}_{i\sigma}c_{i\sigma},
\end{split}
\end{equation}
where $<i,j>$ represents the nearest-neighbor hopping,
$\sigma=\uparrow,\downarrow$ the up and down spin states, and $\ll
i,j\gg$ the next nearest-neighbor hopping. The vector $\mathbf{s}$
are the Pauli matrices. The indices $i,j$ corresponds to site-I or
site-II atoms. The first term of Eq.~\ref{sec:HRS-HRH}) is the
tight-binding Hamiltonian of the honeycomb lattice. The second and
third terms of Eq.~(\ref{sec:HRS-HRH}) are the Haldane and Rashba
Hamiltonians, respectively. The factor $\nu_{ij}$ is the sign of
the current loop $\mathbf{d}_{I,II}\times\mathbf{d}_{II,I}$ or
$\mathbf{d}_{II,I}\times\mathbf{d}_{I,II}$ as shown in the dashed
lines of Fig.~\ref{figLattice}(a). The fourth term of
Eq.~(\ref{sec:HRS-HRH}) is the on-site potential, which can be
experimentally realized \cite{Jo2014}. $\xi_I=+1$ and
$\xi_{II}=-1$ are for site-I and site-II atoms, respectively. The
on-site potential term
$G\sum_{i\sigma}\xi_ic^{\dag}_{i\sigma}c_{i\sigma}$ generates the
band-touching phenomenon and plays the key role for the physics
discussed in this paper [see \ref{sec:SP}].

Using $q_1=\mk\cdot\mathbf{a}_1$ and $q_2=\mk\cdot\mathbf{a}_2$,
Eq.~(\ref{sec:HRS-HRH}) can be written as
\begin{equation}\label{HRmatrix}
H=\left(\begin{array}{cccc}
G+Z_+&X+iY&0&i R_-\\
X-iY&-G+Z_-&-i R^{\ast}_+&0\\
0&i R_+&G+Z_+&X+iY\\
-i R_-^{\ast}&0&X-iY&-G+Z_-\\
\end{array}\right),
\end{equation}
where the wave vector is $(c^{\dag}_{\mk I\uparrow},c^{\dag}_{\mk
II\uparrow},c^{\dag}_{\mk I\downarrow},c^{\dag}_{\mk
II\downarrow})^T$ and
$Z_+=2t_2\left[\cos(q_1-\phi)+\cos(q_2+\phi)+\cos(q_2-q_1-\phi)\right]$,
$Z_-=2t_2\left[\cos(q_1+\phi)+\cos(q_2-\phi)+\cos(q_2-q_1+\phi)\right]$,
$X=t[1+\cos(q_1)+\cos(q_2)], Y=t[\sin(q_1)+\sin(q_2)]$,
$R_-=\alpha(-1+\exp[i(q_1-\frac{\pi}{3})]+\exp[i(q_2+\frac{\pi}{3})])$,
$R_+=\alpha(-1+\exp[i(q_1+\frac{\pi}{3})]+\exp[i(q_2-\frac{\pi}{3})])$.
Using the basis vectors $\mathbf{a}_1$ and $\mathbf{a}_2$ shown in
Fig.~\ref{figLattice}(a), we arrive at the transformation between
the $\mq=(q_1,q_2)$ and $\mk=(k_x,k_y)$ spaces,
\begin{equation}
\begin{split}
&q_1=-(\sqrt{3}/2)k_x+(3/2)k_y,\\
&q_2=(\sqrt{3}/2)k_x+(3/2)k_y.\\
\end{split}
\end{equation}

\begin{table}
\caption{Reciprocal lattice in $\mathbf{k}$ and $\mathbf{q}$ space}
\begin{ruledtabular}
\begin{tabular}{cc}
$\mathbf{k}$-space&$\mathbf{q}$-space\\ \hline
$\mathcal{G}_1=(4\pi/3a)(-\sqrt{3}/2,1/2)$&$\mathcal{G}_1=(2\pi/a)(1,0)$\\
$\mathcal{G}_2=(4\pi/3a)(\sqrt{3}/2,1/2)$&$\mathcal{G}_2=(2\pi/a)(0,1)$\\
\end{tabular}\label{Tab:kqspace}
\end{ruledtabular}
\end{table}
The reciprocal lattice vector $(\mathcal{G}_1,\mathcal{G}_2)$ in
$\mathbf{q}$-space transformed from $\mathbf{k}$-space is given in
Table \ref{Tab:kqspace}. Therefore, we map the honeycomb
reciprocal lattice to a square lattice with the Brillouin zone boundary
$q_1\in(-\pi,\pi)$ and $q_2\in(-\pi,\pi)$.

\begin{figure}
\begin{center}
\includegraphics[width=8cm,height=8cm]{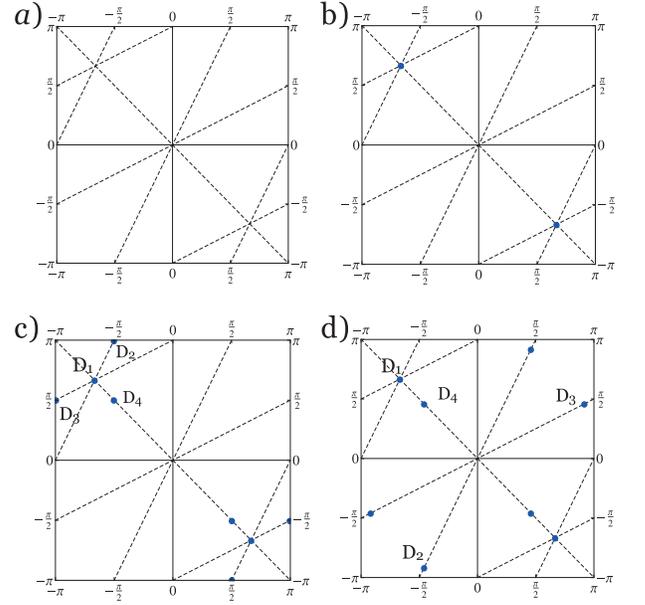}
\end{center}
\caption{The Dirac points (filled circles) in the Brillouin zone.
(a) the Brillouin zone and the dashed line determined by
Eq.~(\ref{DPs}), (b) the two Dirac points in the Haldane system,
(c) the eight Dirac points in the Haldane-Rashba system with
$0<\alpha<1/\sqrt{2}$, and (d) the eight Dirac points in the
Haldane-Rashba system with $\alpha>1/\sqrt{2}$.}\label{figDps}
\end{figure}

The eigenvalue of Eq.~(\ref{HRmatrix}) is given by
\begin{equation}\label{Eigenvalues}
E_{n\sigma}=\frac{1}{2}(Z_-+Z_+)+n\sqrt{\left(G-\frac{Z_--Z_+}{2}\right)^2+\frac{\mathcal{D}_{\sigma}}{2}},
\end{equation}
where $n=\pm$, $\sigma=\pm$, and
\begin{equation}
\begin{split}
\mathcal{D}_{\pm}=&|R_-|^2+|R_+|^2+2|X_0|^2\\
&\pm\sqrt{\left(|R_-|^2-|R_+|^2\right)^2+4|R_-X_0^*-R_+^*X_0|^2},
\end{split}
\end{equation}
with $X_0=X+iY$. We note that $\mathcal{D}_{\pm}$ is independent
of the Haldane hopping interaction $t_2$ and the Haldane phase $\phi$.
The lower of the two conduction bands is $E_{+-}$, and the upper
of the two valence bands is $E_{--}$. The band gap of the
Haldane-Rashba system is thus given by
\begin{equation}
E_g=2\sqrt{\left(G-\frac{Z_--Z_+}{2}\right)^2+\frac{\mathcal{D}_-}{2}}.
\end{equation}
The Chern number changes when the band-touching phenomenon occurs.
In this case, we tune the on-site potential to change the band
gap. It can be shown that $\mathcal{D}_-$ is always positive in
the Brillouin zone and hence, $E_g=0$ leads to the result,
\begin{equation}
G=\frac{Z_--Z_+}{2},~~\mathcal{D}_-=0.
\end{equation}
The condition $\mathcal{D}_-=0$ can be simplified to the result
\begin{equation}\label{DPs}
|R_-R_++X_0^2|^2=0.
\end{equation}
The dashed lines shown in Fig.~\ref{figDps}(a) are obtained from
Eq.~(\ref{DPs}), which provides the possible locations of the
Dirac points. In this paper, we consider only small positive
values of $\alpha$ ($0<\alpha<t$)\cite{Qiao2010,Zarea2009}. In the
absence of a Rashba spin-orbit interaction,
$\mathcal{D}_{\pm}=2|X_0|^2$. The energy gap is then given by
$E_g=\sqrt{(G-(Z_--Z_+)/2)^2+|X_0|^2}$. If we would like to tune
the on-site potential $G$ to construct the band-touching
phenomenon, then $E_g=0$ leads to the solution
$G=(Z_--Z_+)/2+\sqrt{-|X_0|^2}$. Thus, the Dirac points are
determined by $|X_0|^2=0$, which can also be obtained from
Eq.~(\ref{DPs}). The Dirac points for the Haldane system are the
set of $K$ points $(2\pi/3,-2\pi/3)$, $(4\pi/3,2\pi/3)$, and
$(-2\pi/3,-4\pi/3)$ and the set of $K'$ points $(-2\pi/3,2\pi/3)$,
$(2\pi/3,4\pi/3)$, and $(-4\pi/3,-2\pi/3)$. In the Brillouin zone,
the Dirac points in the Haldane system are
$(\pm2\pi/3,\mp2\pi/3)$, as shown in Fig.~\ref{figDps}(b). One
Dirac point is located at $\mathbf{q}$ and the other at
$-\mathbf{q}$. The two Dirac points form a Dirac pair. In the
presence of the Rashba spin-orbit interaction, besides the Dirac
pair
\begin{equation}\label{DPs-D1}
D_1:(-2\pi/3,2\pi/3), D^{\prime}_1:(2\pi/3,-2\pi/3)
\end{equation}
there are six other Dirac points, shown in Fig.~\ref{figDps}(c)
and (d)[see Fig.~\ref{figLattice}(b) in k space]. Furthermore, we
note that when the Rashba interaction is increased, the Dirac
points $D_2$ and $D_3$ (and their Dirac pair partners) approach
the Brillouin zone boundary and appear in the Brillouin zone as
shown in Fig.~\ref{figDps}(c) and (d). The strength of the Rashba
interaction is $\alpha_c=1/\sqrt{2}$. For $\alpha<\alpha_c$, three
of the six Dirac points satisfying Eq.~(\ref{DPs}) are given by
$D_i:(q_1^0,q_2^0)$,
\begin{equation}\label{DPs-D234S}
D_2:(-\theta_0,2(\pi-\theta_0)),~D_3:(2(\theta_0-\pi),\theta_0),~D_4:(-\theta_0,\theta_0),
\end{equation}
where
\begin{equation}
\theta_0=\cos^{-1}\left(\frac{2(\alpha/t)^2-1}{2(\alpha/t)^2+2}\right),
\end{equation}
and the other three Dirac points are given by
$D_i^{\prime}:(-q^0_1,-q^0_2)$. Based on the dashed line
satisfying Eq.~(\ref{DPs}) and shown in Fig.~\ref{figDps}, the eight
Dirac points in the Haldane-Rashba system are shown in Fig.~\ref{figDps}(c).
For $\alpha>\alpha_c$, we have
\begin{equation}\label{DPs-D234L}
D_2:(-\theta_0,-2\theta_0),~D_3:(2\theta_0,\theta_0),~D_4:(-\theta_0,\theta_0).
\end{equation}

In the next section, we will obtain the Chern number pattern of
the Haldane-Rashba system by investigating the band-touching
phenomenon of these eight Dirac points. For convenience, we scale
the energy by the hopping energy $t$, i.e., we set $t=1$ in the
following calculations.


\section{Chern number pattern and edge states}\label{sec:Ch pattern}
In this section, we will obtain the Chern number pattern of the
Haldane-Rashba system. First, we construct an effective two-band
model from the Hamiltonian Eq.~(\ref{HRmatrix}) to map the
$\mathbf{q}$ space to the $\mathbf{d}(\mathbf{q})$ space and calculate
the Pontryagin winding number wrapped by the $\mathbf{d}$ vector. We
also use the Berry curvature method to present the numerical result of
the Chern number pattern. The resulting edge states are also discussed
in this section.


Equation (\ref{DPs}) determines the locations of the Dirac points.
This implies that the four-band Haldane-Rashba Hamiltonian can be
cast as an effective two-band model for determining the Chern
number pattern
\begin{equation}
H_{\mathrm{eff}}=\left(\begin{array}{cc}
d_z&d_x-id_y\\
d_x+id_y&-d_z
\end{array}\right).
\end{equation}
The Chern number for the two-band model can be determined by the
Pontryagin winding number method \cite{Hsiang2001, Sticlet2012}
\begin{equation}\label{Ch}
\mathrm{Ch}=\frac{1}{2}\sum_{\mathbf{q}\in
D_i,D_i^{\prime}}\sgn(\frac{\partial\mathbf{d}}{\partial
q_1}\times\frac{\partial\mathbf{d}}{\partial q_2})_z\sgn(d_z),
\end{equation}
where $\sgn(\cdots)$ indicates the sign of $(\cdots)$, $z$ an
arbitrary axis selected in this pseudo-spin space, and $D_i$ a set
of Dirac points in the Brillouin zone \footnote{The overall sign
of the Chern number in Eq.~(\ref{Ch}) depends on the Jacobian
transformation and the overall sign chosen in $d_y$ and $d_x$.
However, the overall sign choice does not alter the conclusion of
this paper. The Jacobian transformation from $(k_x,k_y)$ to
$(q_1,q_2)$ is $+1$, and thus, in this paper, we choose the same
sign convention for $d_x$ and $d_y$ as in
Ref.~\cite{Haldane1988}}. The two terms $d_x$ and $d_y$ determine
the positions of the Dirac points by the requirements $d_x=0$ and
$d_y=0$. Equation (\ref{DPs}) is equivalent to the vanishing real
as well as imaginary part of $R_-R_++X_0^2$. Because
Eq.~(\ref{Ch}) requires only the sign of each term in the bracket,
the terms $d_x$, $d_y$, and $d_z$ accounting for Eq.~(\ref{DPs})
can be chosen as
\begin{equation}\label{d-vec}
\begin{split}
&d_x=\sqrt{\tilde{d}_x},~\tilde{d}_x=\Re[R_-R_++X_0^2],\\
&d_y=\sqrt{\tilde{d}_y},~\tilde{d}_y=\Im[R_-R_++X_0^2],\\ 
&d_z=G-(Z_--Z_+)/2.\\
\end{split}
\end{equation}
Substituting Eqs.~(\ref{DPs-D1}) and (\ref{DPs-D234S})
[or Eqs.~(\ref{DPs-D1}) and (\ref{DPs-D234L})] in $d_z$, we derive
\begin{equation}\label{sign-dz}
\begin{split}
&d_z(D_1)=G-G_e\sin\phi ,~d_z(D_{2,3,4})=G-M_{\alpha}\sin\phi\\
&d_z(D_1^{\prime})=G+G_e\sin\phi,~d_z(D^{\prime}_{2,3,4})=G+M_{\alpha}\sin\phi,
\end{split}
\end{equation}
where $M_{\alpha}= G_eG_{\alpha}$ and (note that $t=1$ was used)
\begin{equation}\label{ChHR2}
G_e=3\sqrt{3}t_2,~G_{\alpha}=\frac{\sqrt{1+4\alpha^2}}{(1+\alpha^2)^2}.
\end{equation}
On the other hand, we note that $R_-R_++X_0^2$ vanishes at the Dirac
points. The ambiguity appearing in the denominator after a partial
derivative with respect to $\mq$ should be neglected in the sign function. The
sign of the term
$(\partial_{q_1}\mathbf{d}\times\partial_{q_2}\mathbf{d})_z$ can
be determined by
$(\partial_{q_1}\tilde{\mathbf{d}}\times\partial_{q_2}\tilde{\mathbf{d}})_z$,
where $\tilde{\mathbf{d}}=(\tilde{d}_x,\tilde{d}_y)$ [see Eq.~(\ref{d-vec})].
Substituting Eqs.~(\ref{DPs-D1}) and (\ref{DPs-D234S}) [or Eqs.~(\ref{DPs-D1}) and (\ref{DPs-D234L})] in
$J_z\equiv(\partial_{q_1}\tilde{\mathbf{d}}\times\partial_{q_2}\tilde{\mathbf{d}})_z$,
we have
\begin{equation}\label{sign-Jz}
\begin{split}
&J_z(D_1)=(9\sqrt{3}/2)\alpha^4 ,~J_z(D_{2,3,4})=-(27\sqrt{3}/2)\alpha^4G_{\alpha}\\
&J_z(D_1^{\prime})=-(9\sqrt{3}/2)\alpha^4,~J_z(D^{\prime}_{2,3,4})=(27\sqrt{3}/2)\alpha^4G_{\alpha}.
\end{split}
\end{equation}
Using Eqs.~(\ref{sign-dz}) and (\ref{sign-Jz}), we can
calculate the Chern number in each zone enclosed by the phase
boundary. For example, considering $0<\phi<\pi$, the result is given
in Table \ref{Tab:ChHR1}.

\begin{table}
\caption{Values of the Chern number Ch in a Haldane-Rashba system with $0<\phi<\pi$.}
\begin{ruledtabular}

\begin{tabular}{cccccc}
Dirac points&$D_1$&$D_{2,3,4}$&$D_1^{\prime}$&$D_{2,3,4}^{\prime}$&Ch\\
Mass&$G-G_e$&$G-M_{\alpha}$&$G+G_e$&$G+M_{\alpha}$&\\
$\sgn(J_z)$&$+$&$-$&$-$&$+$&\\ \hline
$\sgn(J_z)\sgn(d_z)$\\
$G<-G_e$&$-$&$+$&$+$&$-$&0\\
$-G_e<G<-M_{\alpha}$&$-$&$+$&$-$&$-$&$-1$\\
$-M_{\alpha}<G<M_{\alpha}$&$-$&$+$&$-$&$+$&$+2$\\
$M_{\alpha}<G<G_e$&$-$&$-$&$-$&$+$&$-1$\\
$G_e<G$&$+$&$-$&$-$&$+$&0\\

\end{tabular}\label{Tab:ChHR1}

\end{ruledtabular}
\end{table}

\begin{figure}
\begin{center}
\includegraphics[width=8cm,height=8cm]{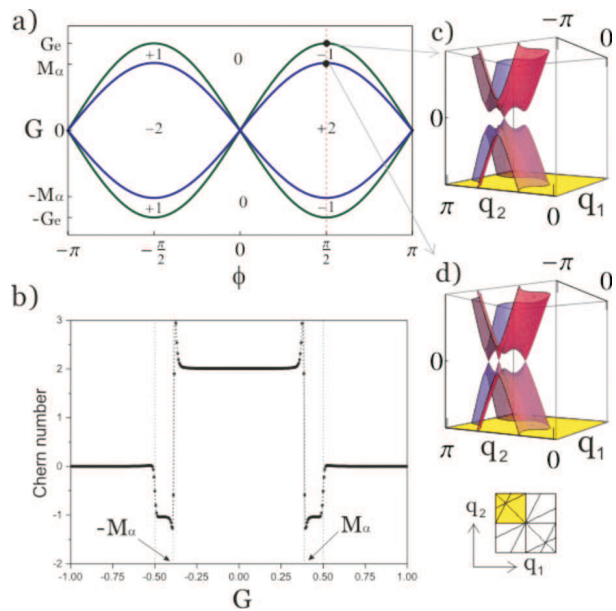}
\end{center}
\caption{(Color online) (a) The phase diagram and (b) the
numerical result of the Chern number pattern of the Haldane-Rashba
system. The band structures at the phase boundary is shown in (c)
and (d). We have drawn the band structure at only one quadrant of
the Brillouin zone to show the occurrence of the band-touching
phenomenon. The band structure at the opposite quadrant is fully
gapped.}\label{figHRCh}
\end{figure}

Note that $\alpha^4$ and $G_{\alpha}$ are always positive, and the
sign of $J_z$ can easily be obtained from Eq.~(\ref{sign-Jz}) at
each Dirac point. Substituting Eqs.~(\ref{sign-dz}) and
(\ref{sign-Jz}) into Eq.~(\ref{Ch}) and summing over these eight
Dirac points, we obtain
\begin{equation}\label{ChHR1}
\begin{split}
\mathrm{Ch}=&\frac{1}{2}\left[\sgn\left(G-G_e\sin\phi\right)-\sgn\left(G+G_e\sin\phi\right)\right]\\
&+\frac{3}{2}\left[\sgn\left(G+M_{\alpha}\sin\phi\right)-\sgn\left(G-M_{\alpha}\sin\phi\right)\right].
\end{split}
\end{equation}
Therefore, the phase diagram of the Haldane-Rashba system can be
determined by Eqs.~(\ref{ChHR1}) and (\ref{ChHR2}). The result is
shown in Fig.~\ref{figHRCh}(a). The Haldane-Rashba system exhibits
the Chern number pattern $\{+2,-1,0\}$ and $\{-2,+1,0\}$ for a
positive and negative Haldane phase, respectively. A system state
with a non-zero Chern number is called a Chern-insulating state.
There are two boundaries exhibited by the two parameters
$M_{\alpha}$ and $G_e$. Furthermore, the change in Chern number is
$\Delta\mathrm{Ch}=\pm N$, where $N$ is the number of Dirac points
occurring simultaneously in a band-touching
phenomenon~\cite{Haldane1988, Murakami2008}. It is interesting to
note that when the Chern number changes from $+2$ to $-1$ (or $-2$
to $+1$), the band-touching phenomenon occurs at the three Dirac
points $D_2, D_3$, and $D_4$ simultaneously, as shown in
Fig.~\ref{figHRCh}(c) and (d) and theoretically confirmed in Table
\ref{Tab:ChHR1}. The simultaneous band touching leads to the
resulting coefficient $3/2$ in the second square bracket of
Eq.~(\ref{ChHR1}), which is three times larger than the band
touching caused by $G_e$. Interestingly, it was recently found
that the shape of the Dirac cone is related to the spin connection
that couples to the sublattice pseudospin \cite{Yang2015}. On the
other hand, when $\alpha=0$, the Chern number becomes
$\mathrm{Ch}=\sgn(G+G_e\sin\phi)-\sgn(G-G_e\sin\phi)$, which is
doubled because the electron spin is considered. When $t_2=0$,
Eq.~(\ref{ChHR1}) gives $G_e=0$ and $M_{\alpha}=0$, and thus,
$\mathrm{Ch}=0$, which is in agreement with the fact that the
Chern number vanishes in the time-reversal symmetric system.

In order to compare the present result obtained in $\mq$ space, we
also calculate the Chern number numerically by using the Berry
curvature method in $\mk$ space,
\begin{equation}
\mathrm{Ch}=-\frac{1}{2\pi}\sum_n\int_{BZ}\Omega^{(n)}_{xy}(\mk)dk_xdk_y,
\end{equation}
where
\begin{equation}
\Omega^{(n)}_{xy}=\sum_{n'(\neq
n)}\frac{2\Im\langle\psi_{nk}|\frac{\partial H}{\partial
k_x}|\psi_{n'k}\rangle\langle\psi_{n'k}|\frac{\partial H}{\partial
k_y}|\psi_{nk}\rangle}{(E_{n'}(\mk)-E_{n}(\mk))^2}.
\end{equation}
We consider the Haldane-Rashba system with $\phi=\pi/2$,
$\alpha=0.7$, and $t_2=1/(6\sqrt{3})$. We have $G_e=0.5$ and
$M_{\alpha}\simeq0.38$. The result is shown in Fig.~\ref{figHRCh}
(b), which is in agreement with our Pontryagin winding number
method.

We also numerically calculate the upper one of two valence bands
and the lower one of the two conduction bands in order to check
the band-touching phenomenon. In the change of total Chern number
from $-1$ to $+2$, we find that the upper one of two valence bands
($\mathrm{Ch}=-1$) increases by $3$ and becomes $+2$, and the
lower one of two conduction bands ($\mathrm{Ch}=+1$) decreases by
$3$ and becomes $-2$. The Chern number of the lower one of two
valence bands is numerically found to be always zero, and the
total Chern number of the two valence bands indeed changes from
$-1$ to $+2$.

\begin{figure*}
\begin{center}
\includegraphics[width=16cm,height=7cm]{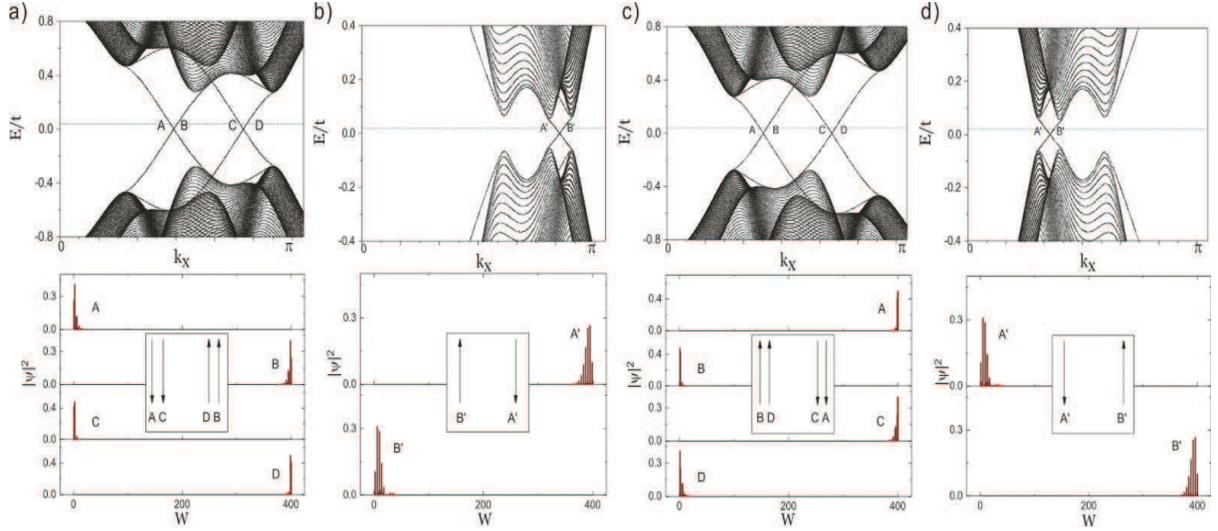}
\end{center}
\caption{Edge state and the corresponding wave function
distribution and the current direction of the Haldane-Rashba
system with $\alpha=0.7$, $t_2=1/(6\sqrt{3})$, and $E_F=0.01$ for
(a) $G=0.1t$, (b) $G=0.45t$ at $\phi=+\pi/2$, (c) $G=0.1t$, and (d)
$G=0.45t$ at $\phi=-\pi/2$. The ribbon denoted $W$ is 400 atoms wide.
}\label{EdgeWF}
\end{figure*}

Next, we turn to the discussion of the resulting edge states in the
Haldane-Rashba system. We consider the strength of the Rashba
spin-orbit interaction $\alpha=0.7$ ($M_{\alpha}=0.38$),
$t_2=1/(6\sqrt{3})$ ($G_e=0.5$), and fix the two Haldane phases at
$\phi=\pi/2$ and $\phi=-\pi/2$. Tuning the on-site potential $G$,
we get the Chern number pattern $\{+2,-1,0\}$ for the positive
Haldane phase $\phi=+\pi/2$ for which the phase transition occurs
at $G=\pm M_{\alpha}$ [obtained from Eq. (\ref{ChHR2})] for
$\{+2,-1\}$ and $G=\pm G_e$ for $\{-1,0\}$. The edge states and
the corresponding wave function distribution $|\psi|^2$ for the Chern
numbers $+2$ and $-1$ are shown in Fig.~\ref{EdgeWF}(a) and (b).

Figure \ref{EdgeWF}(a) and (b) show the edge states and their
wave function distributions at the positive Haldane phase
$\phi=\pi/2$, which is located explicitly at the edge of the honeycomb
ribbon. Based on the slope of each edge state in the band
structure, the current flow could be obtained. The edge states
$D,B$ are located at the right hand side of the ribbon and $A,C$ at the left
hand side. The Chern number of Fig.~\ref{EdgeWF}(a) is thus
identified as $+2$.  When the on-site potential changes from
$G=0.1$ to $G=0.45$ passing through $G=0.38$, a phase transition
occurs and the edge state $BD$ ($AC$) reduces to the edge state $A'$
($B'$) as shown in Fig.~\ref{EdgeWF}(b). There the Chern number
now becomes $-1$. For the negative Haldane phase $\phi=-\pi/2$ [see
Fig.~\ref{EdgeWF}(c) and (d)], the edge states $AC$ and $B'$ are located
at the right hand side and $BD$ and $A'$ at the left hand side
of the ribbon, respectively.

Thus, we have obtained the Chern number pattern and determined
the regime of each Chern-insulating phase of the Haldane-Rashba
system. We used the Pontryagin winding number method in $\mathbf{q}$
space to determine the Chern number and the phase boundary, and
the result is in agreement with the numerical method calculated in
$\mathbf{k}$ space. In the next section, we study the
variation of the spin-polarization of the edge states across the phase
boundary.

\section{Spin polarization of edge states}\label{sec:SP}
Because the edge current is non-zero in the Chern insulating
states, the interplay of orbital motion (current loop formed by
the edge currents) and the Rashba spin-orbit interaction may cause
a non-zero out-of-plane spin polarization.  We note that the
zigzag graphene ribbon naturally form site-I atoms (more neighbors
of site-II atoms) at the right hand and site-II atoms (more
neighbors of site-I atoms) at the left hand side of the ribbon,
respectively [see Fig.~\ref{figLattice}]. In this sense, the two
edges of the ribbon must have different Haldane current loops to
exhibit an edge current, as shown in Fig.~\ref{figLattice}(a). On
the other hand, the Rashba spin-orbit interaction indicates that
the in-plane spin $\mathbf{s}$ interacts with the nearest-neighbor
hopping vector $\mathbf{d}_{ij}$. Because the two edges of the
ribbon have opposite hopping directions between site-I and site-II
atoms, the cross product of the in-plane spin and nearest-neighbor
hopping vectors implies that the sign of spin polarization on one
side is opposite to that on the other side.

The Rashba spin-orbit coupling governs the nearest neighbor
hopping of different site atoms. At right (left) side of the
ribbon, the site-I (site-II) atom has more nearest neighbors than
site-II (site-I) atoms. Therefore, site-I (site-II) atoms
contributes more Rashba spin-orbit hopping to the site-II atoms at
right (left) hand side of ribbon. Therefore, the spin polarization
of the edge states at the right (left) hand side of the ribbon
would be dominated by site-II (site-I) atoms.

Furthermore, we note that the on-site potential at site-II is
negative compare to that at site-I. The effective Hamiltonian for
the edge states would be $Gc^{\dag}_{I,\sigma}c_{I,\sigma}$ for
the left hand side and $-Gc^{\dag}_{II,-\sigma}c_{II,-\sigma}$ for
the right hand side of the ribbon, where $-\sigma$ means the
opposite spin states with respect to $\sigma$. When the on-site
potential is turned on, the interplay of the Haldane phase and the
on-site potential would cause the site-I and site-II atoms to
contribute unequally to the out-of-plane spin polarization of the
edge currents.

\begin{figure}
\begin{center}
\includegraphics[width=8cm,height=12cm]{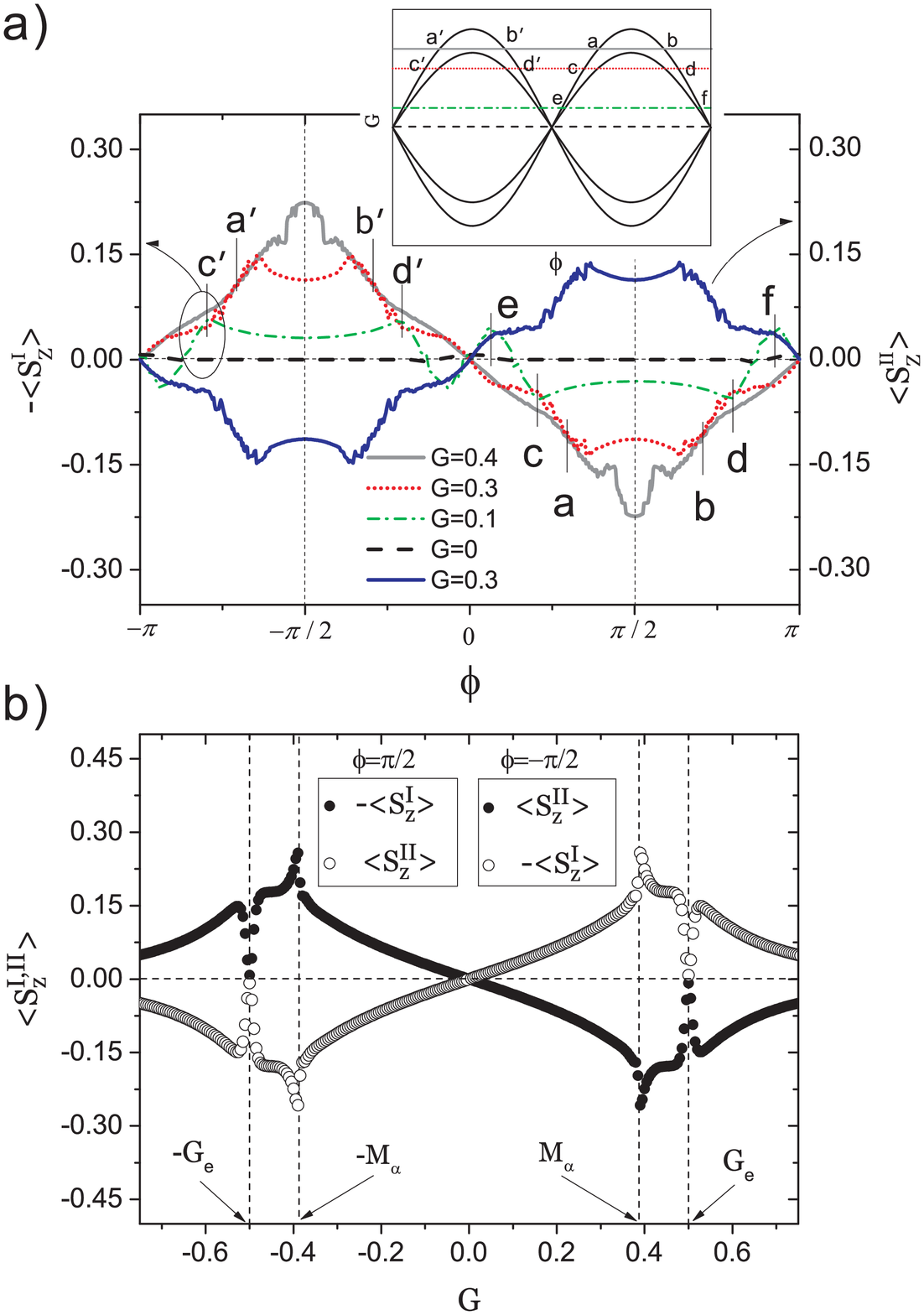}
\end{center}
\caption{(Color online) Linear response of $-S^I_z$ and $S^{II}_z$
to an applied electric field. $\alpha=0.7$ and
$t_2=1/(6\sqrt{3})$. (a) Various Haldane phases and a discrete
on-site potential and (b) various on-site potentials and a
positive Haldane phase $\phi=\pi/2$ ($-\langle
S^I_z\rangle\rightarrow\bullet$, $\langle
S^{II}_z\rangle\rightarrow\circ$) and a negative Haldane phase
$\phi=-\pi/2$ ($-\langle S^I_z\rangle\rightarrow\circ$, $\langle
S^{II}_z\rangle\rightarrow\bullet$).}\label{figBS}
\end{figure}

Interestingly, this also indicates that the on-site potential
behaves like an effective magnetic field for the edge currents. In
this sense, when the sign of the on-site potential is changed, the
spin polarization also changes its sign regardless of the change
in the Chern number. Furthermore, this also implies that the
magnitude of the spin polarization at both edges should also grow
with the on-site potential. Consequently, the spin-polarization
would persist under a change in the Chern number. In order to show
these bulk and edge effects, we first study the bulk spin
polarization and then compare the result to that of the edge
currents. The spin matrices for site-I and site-II are given by
$S_z^{I}$ and $S^{II}_z$, respectively,
\begin{equation}
S^I_z=\left(\begin{array}{cccc}
1&0&0&0\\
0&0&0&0\\
0&0&-1&0\\
0&0&0&0\\
\end{array}\right),~S^{II}_z=\left(\begin{array}{cccc}
0&0&0&0\\
0&1&0&0\\
0&0&0&0\\
0&0&0&-1\\
\end{array}\right).
\end{equation}
The on-site potential would effectively couple to the spins of the
edge states at site-I and -II with $-S^I_z$ and $ S^{II}_z$,
respectively. The linear responses of spin
$\Sigma_z^{I,II}=(\hbar/2)S_z^{I,II}$ to an applied electric field
in the $y$ direction $E_y$ is denoted as $\langle
\Sigma_z^{I,II}\rangle=(e\hbar/8\pi)\langle S^{I,II}_z\rangle
E_y$, where
\begin{equation}\label{LRspin}
\begin{split}
\langle S^{I,II}_z\rangle=&\frac{1}{\pi}\int_{BZ}
dk_xdk_y\\
&\times\sum_{n\neq
n'}\frac{\Im\langle\psi_{nk}|S^{I,II}_z|\psi_{n'k}\rangle\langle\psi_{n'k}|\frac{\partial
H}{\partial k_y}|\psi_{nk}\rangle}{(E_{n'}(\mk)-E_{n}(\mk))^2}.
\end{split}
\end{equation}
$\langle S_z^I\rangle$ and $\langle S_z^{II}\rangle$ in Eq.
(\ref{LRspin}) are calculated numerically and shown in
Fig.~\ref{figBS} where the Rashba spin-orbit interaction
$\alpha=0.7$.
When the on-site potential is fixed and non-zero, the change in
the Haldane phase changes the magnitude of the bulk spin
polarization. When the sign of the Haldane phase is changed, the
sign of the spin polarization also changes because of the Rashba
spin-orbit interaction, as shown in Fig.~\ref{figBS}(a). The inset
in Fig.~\ref{figBS}(a) indicates the outer boundary points for the
change in Chern number. In Fig.~\ref{figBS}(a), we only show the
effect of on-site potential for $-\langle S^I_z\rangle$. The
result for $\langle S^{II}_z\rangle$ can be obtained by replacing
$\phi\rightarrow-\phi$ and $G\rightarrow-G$ as can be seen in
Eq.~(\ref{HRmatrix}). When varying the on-site potential, the
number of edge states and the current direction would change.

The sign of the bulk spin polarization is approximately determined
by the sign of the Haldane phase in the regimes of non-zero Chern
number. For instance, when $\phi>0$ ($\phi<0$), $-\langle
S^I_z\rangle$ is negative (positive) and, as shown in the
intervals $ab$ ($\phi>0$) and $a'b'$ ($\phi<0$) and the intervals
$cd$ ($\phi>0$) and $c'd'$ ($\phi<0$). However, when $G$ is small
and $\phi$ is close to $0$ and $\pi$, the sign of the bulk spin
polarization can not be determined only by the Haldane phase. For
the interval $ef$, the sign of the bulk spin polarization is
altered by the on-site potential, as shown in the interval $ec$
and $df$ for $G=0.1$, where $\phi$ is close to $0$ and $\pi$. This
is because the Haldane hopping term
$t_2e^{i\nu_{ij}\phi}c^{\dag}_{i\sigma}c_{i\sigma}$ has the same
behavior (effective magnetic field) as the on-site term
$G\xi_ic^{\dag}_{i\sigma}c_{i\sigma}$\footnote{This can be seen as
follows. By eliminating $(Z_++Z_-)/2$ [see
Eq.~(\ref{Eigenvalues})] in the Hamiltonian Eq.~(\ref{HRmatrix}),
the diagonal term becomes $\pm[G-(Z_--Z_+)/2]$ and $Z_--Z_+$ is
proportional to $\sin\phi$, which is an odd function in $\phi$.
Therefore, the Haldane phase behaves like an effective magnetic
field.}.  For small $G$, the competition of the Haldane hopping
term and the on-site term becomes explicit when $\phi$ is close to
$0$ and $\pi$.

In Fig.~\ref{figBS}(b), we first consider the Haldane phase fixed
at $\phi=\pi/2$ and vary the on-site potential for $-\langle
S^I_z\rangle$ and $\langle S_z^{II}\rangle$. The result for
negative Haldane phase $\phi=-\pi/2$ is obtained by just making the
replacement $-\langle S^I_z\rangle\leftrightarrow\langle
S^{II}_z\rangle$. In the interval $-M_{\alpha}<G<M_{\alpha}$, the
Chern number is $+2$. However, the bulk spin polarization changes
sign and magnitude with the on-site potential. In the interval
$M_{\alpha}<G<G_e$, the Chern number is $-1$. Interestingly, the
bulk spin-polarization keeps its sign. That is, near
$\phi=\pm\pi/2$, both the site-I and site-II bulk spin polarization
indeed vary with the on-site potential, and keep their signs even when
the Chern number has changed. To see the bulk-edge
correspondence explicitly for spin-polarization, we will in the following focus
only on the Haldane phases $\phi=\pi/2$ and $\phi=-\pi/2$,
which have the largest phase regimes.

\begin{figure}
\begin{center}
\includegraphics[width=9cm,height=12cm]{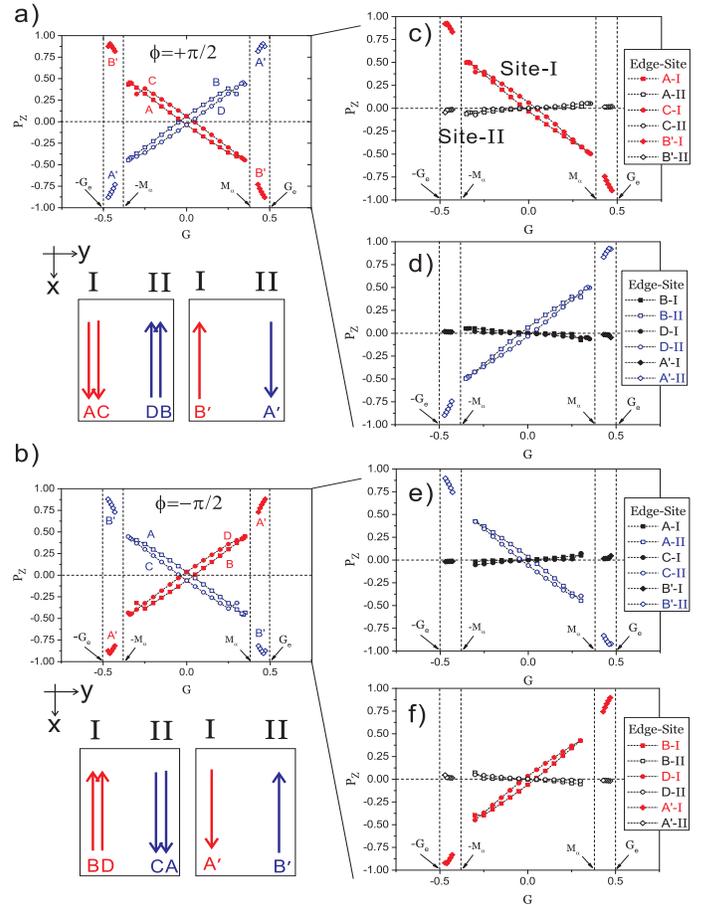}
\end{center}
\caption{ (color online) Numerical results ($\alpha=0.7$ and $
t_2=1/(6\sqrt{3})$) for the change in spin polarization of the
edge states for (a) $\phi=\pi/2$ and (b) $\phi=-\pi/2$ at
$E_F=0.01$. Filled and red (empty and blue) circle or square
represent the edge states at left (right) hand side of ribbon.
Panels (c) and (d) show the contributions of different sites
(site-I and site-II) to the spin polarization of the edge states
in (a). Panels (c) and (d) show that the large component for $A,
C, B'$ is site-I. However, the large component for $B, D, A'$ is
site-II.}\label{figEBSP}
\end{figure}

To confirm the above two distinct results deduced from the
tight-binding Hamiltonian, we calculate the spin polarization of
each edge current versus the on-site potential at the two
different Haldane phases ($\phi=\pi/2$ and $\phi=-\pi/2$). The
wave function is
$\psi=(\psi_{I1\uparrow},\psi_{I1\downarrow},\psi_{II1\uparrow},\psi_{II1\downarrow},\psi_{I2\uparrow},\psi_{I2\downarrow},\cdots)^T$
and the spin polarization $P_z$ is given by
$P_z=|\psi_{I1\uparrow}|^2-|\psi_{I1\downarrow}|^2+|\psi_{II1\uparrow}|^2-|\psi_{II1\downarrow}|^2+\cdots$.
The numerical results for the spin polarization of the edge
currents for the positive ($\phi=+\pi/2$) and negative
($\phi=-\pi/2$) Haldane phases are shown in Fig.~\ref{figEBSP}(a)
and (b), respectively.

Consider the positive Haldane phase [see Fig.~\ref{figEBSP}(a)].
As the on-site potential increases from $-M_{\alpha}$ to
$M_{\alpha}$, the spin polarization of the edge currents $B$ and $D$ increases.
However, the spin polarization of the edge states $A$ and $C$ decreases.
Interestingly, as the on-site potential passes
through $M_{\alpha}$ and $-M_{\alpha}$, the direction of the
spin polarization does not change at each edge of the ribbon,
even though the chirality of the Chern number has changed from
$+2$ to $-1$. This phenomenon is the same for the negative Haldane
phase shown in Fig.~\ref{figBS}(b).

Importantly, compare Figs.~\ref{figBS}(b) and
\ref{figEBSP}(a) at $\phi=\pi/2$. We find that the edge states $B,
D, A'$ behave like $\langle S^{II}_z\rangle$ and that the edge states
$A, C, B'$ behave like $-\langle S^{I}_z\rangle$ for a positive
Haldane phase. The contributions of site-I and site-II to the spin
polarization of the edge currents at $\phi=\pi/2$ are shown in
Fig.~\ref{figEBSP}(c) and (d).
Here, the spin polarization is calculated only for one site, i.e.,
$P_z=|\psi_{I1\uparrow}|^2-|\psi_{I1\downarrow}|^2+|\psi_{I2\uparrow}|^2-|\psi_{I2\downarrow}|^2+\cdots$
for site-I and
$P_z=|\psi_{II1\uparrow}|^2-|\psi_{II1\downarrow}|^2+|\psi_{II2\uparrow}|^2-|\psi_{II2\downarrow}|^2+\cdots$
for site-II. Figure \ref{figEBSP}(c) shows that the spin
polarization of edge states $A, C$, and $B'$ at site-$I$ is larger
than at site-$II$. Figure \ref{figEBSP}(d) shows that at a
positive Haldane phase, the spin polarization of $B, D$, and $A'$
at site-II is indeed larger than that at site-I. We find that the
spin polarization at site-II (site-I) is the large component
and at site-I (site-II) is the small component for the edge states
$B, D, A'$ ($A, C, B'$) for a positive Haldane phase. On the other
hand, we also find that for a negative Haldane phase
[see Figs.~\ref{figEBSP}(b) and \ref{figBS}(b)] the spin
polarization of site-II is the large component for $A, C, B'$
[see Fig.~\ref{figEBSP}(e)], and the spin polarization of
site-I is the large component for the edge states $B, D, A'$
[see Fig.~\ref{figEBSP}(e)].

The spin polarization of the edge current at the right hand side of the
ribbon [compare to Fig.~\ref{figLattice}] has a large contribution
from site-II, and the opposite side of the ribbon has a large
contribution from site-I. Furthermore, the sign of the spin
polarization of the edge current is the same as the bulk spin
polarization and is determined by the on-site potential, which is the
approximate bulk-edge correspondence of spin polarization. We find
that the spin polarization of the edge currents indeed persists
regardless of the change in the Chern number, because the
on-site potential behaves like an effective magnetic
field and the Haldane current loops contribute unequally to the
edges of the ribbon.



\begin{figure}
\begin{center}
\includegraphics[width=8cm,height=5cm]{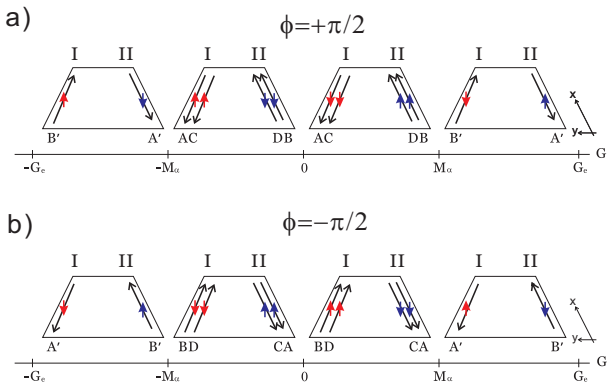}
\end{center}
\caption{(Color online) Schematic diagram showing the edge state
current distribution (long arrows) and spin polarization for (a) a
positive Haldane phase $\phi=\pi/2$ and (b) a negative Haldane phase
$\phi=-\pi/2$.}\label{figSC}
\end{figure}

The spin polarization [Fig.~\ref{figEBSP}] and current
distribution [Fig.~\ref{EdgeWF}] for varying Haldane phase and
on-site potential in real space is schematically summarized in
Fig.~\ref{figSC}(a) and (b).

If the on-site potential is fixed, a change in the Haldane phase
leads to a change in the sign of the spin polarization and the direction
of the edge current. However, if the Haldane phase is fixed, a
change in the sign of the on-site potential leads only to a change in
the sign of the spin polarization. Once the sign of the on-site potential
is fixed, the spin polarization persists under a change in the Chern
number. In this sense, the bulk-protected spin polarization of
the edge states in the Haldane-Rashba system can be controlled in a
Chern-insulating phase. Furthermore, by varying the sign of the
Haldane phase and the magnitude of the on-site potential, the
current can be reduced without changing the polarization of spin.

The present result is rather different from the Kane-Mele-Rashba
system \cite{Chen2011} where the spin polarization in quantum
spin-Hall phase persists in the presence of an exchange field in
zero Chern number regime. Here, we study the spin polarization in
non-zero Chern number regimes. The Haldane orbital Hamiltonian in
Ref. \cite{Chen2011} describes the coupling of the orbital angular
momentum and the exchange field in which $\nu_{ij}$ behaves like
an effective orbital angular momentum and does not cause the
Haldane phase $\phi$. In this paper, the Haldane phase $\phi$
plays an important role in changing the spin polarization and
non-zero Chern number regimes. We emphasize that in this paper the
orientation of spin polarization of edge currents can be
controlled in a single and non-zero Chern-insulating phase.

In short, the Haldane current loop naturally determines the
contribution of site atoms in a zigzag ribbon to the spin
polarization of the edge currents. The on-site potential plays the
role of an effective magnetic field, and the variation of the spin
polarization of the edge state approximately obeys the bulk spin
polarization of the site atoms near $\phi=\pm\pi/2$, which forms
the approximate bulk-edge correspondence to the spin polarization.
When $\phi$ approaches $0$ and $\pm\pi$, the competition of the
on-site potential and Haldane hopping term may break the
correspondence. Because of the bulk protected spin polarization,
the orientation of the spin polarization can be manipulated in a
single and non-zero Chern-insulating phase.

\section{Conclusion}\label{sec:Conclusion}
We have investigated the Chern number pattern of the
Haldane-Rashba system by using the Pontryagin winding number and
numerical Berry curvature methods. We found that the on-site
potential is the generator of the band-touching phenomenon and can
be used to change the Chern number of the system. Three Chern
insulating regimes can be exhibited by tuning the on-site
potential. The Chern number pattern is $\{+2,-1,0\}$ for positive
Haldane phase and $\{-2,+1,0\}$ for negative Haldane phase. If the
on-site potential is fixed, the change in sign of the Haldane
phase changes the direction of the edge currents and the sign of
the out-of-plane spin polarization. Interestingly, we found that
the spin polarization of the edge states persists under the change
in the Chern number. Furthermore, the spin polarization of the
edge states vary with the on-site potential, which implies that
the on-site potential effectively couples to the spin of the edge
states and behaves like an effective magnetic field. We calculated
the bulk spin polarization and found that the spin polarization of
the edge states approximately follows the bulk spin polarization.
This indicates that the zigzag ribbon with two different Haldane
current loops exhibits an approximate bulk-edge correspondence to
the spin polarization. We found that the orientation of the spin
polarization of the edge states could vary in a single
Chern-insulating phase by tuning the on-site potential, which
would play a crucial role in spintronics.

\section*{ACKNOWLEDGMENTS}
T. W. C. would like to thank S. Murakami for valuable discussions
regarding the Chern number and W.-X. Feng for the topological
insulator. T. W. C and C. D. H thank the Taiwan Ministry of
Science and Technology for financial support under Grant No.\ MOST
104-2112-M-110-003 and NSC 101-2112-M-002-016-MY3.



\begin{thebibliography}{99}
\bibitem{Berry1984} M. Berry, Proc.\ Royal Soc.\ London, Ser.\ A {\bf
392}, 45 (1984).
\bibitem{Thouless1982} D. J. Thouless, M. Kohmoto, M. P.
Nightingale, and M. den Nijs, Phys.\ Rev.\ Lett.\ {\bf 49}, 405
(1982). S. S. Chern, Ann.\ Math.\ {\bf 47}, 85 (1946).

\bibitem{Hatsugai1993} Y. Hatsugai, Phys.\ Rev.\ Lett. {\bf 71},
3697 (1993); Y. Hatsugai, Phys.\ Rev.\ B {\bf 48}, 11851 (1993);
H. Watanabe, Y. Hatsugai, and H. Aoki, Phys.\ Rev.\ B {\bf 82},
241403(R) (2010).

\bibitem{Haldane1988} F. D. M. Haldane, Phys.\ Rev.\ Lett.\ {\bf 61},
2015 (1988).

\bibitem{Kane2005} C. L. Kane and E. J. Mele, Phys.\ Rev.\ Lett.\ {\bf
95}, 2266801 (2005); C. Xu and J. E. Moore, Phys.\ Rev.\ B {\bf
73}, 045322 (2006).

\bibitem{Bernevig2006} B. A. Bernevig, T. L. Hughes, and S.-C. Zhang, Science {\bf 314}, 1757 (2006); M. Konig {\it et al.}, Science {\bf 318}, 766
(2007).

\bibitem{Fu2007} L. Fu, C. L. Kane, and E. J. Mele, Phys.\ Rev.\
Lett.\ {\bf 98}, 106803 (2007); J. E. Moore and Balents, Phys.\ Rev.\
B {\bf 75}, 121306 (2007); R. Roy, Phys.\ Rev.\ B {\bf 79}, 195322
(2009); L. Fu and C. L. Kane, Phys.\ Rev.\ B {\bf 76}, 045302
(2007).

\bibitem{Hsieh2008} D. Hsieh, D. Qian, L. Wray, Y. Xia, Y. S. Hor,
R. J. Cava, and M. Z. Hasan, Nature (London) {\bf 452}, 970
(2008).


\bibitem{Hasan2010} M. Z. Hasan and C. L. Kane, Rev.\ Mod.\ Phys.\ {\bf
82}, 3045 (2010); X.-L. Qi and S.-C. Zhang, Rev.\ Mod.\ Phys.\ {\bf
83}, 1057 (2011).

\bibitem{Sheng2006} D. N. Sheng, Z. Y. Weng, L. Sheng, and F. D.
M. Haldane, Phys.\ Rev.\ Lett.\ {\bf 97}, 036808 (2006).

\bibitem{Prodan2009} E. Prodan,
Phys.\ Rev.\ B {\bf 80}, 125327 (2009); E. Prodan, New J. Phys.\ {\bf
12}, 065003 (2010). H. Li, L. Sheng, D. N. Sheng, and D. Y. Xing,
Phys.\ Rev.\ B {\bf 82}, 165104 (2010).

\bibitem{Shan2010} W.-Y. Shan, H.-Z. Lu, and S.-Q. Shen, New J. Phys.\
{\bf 12}, 043048 (2010); T. Fukui and Y. Hatsugai, Phys.\ Rev.\ B
{\bf 75}, 121403 (2007); A. M. Essin and J. E. Moore, Phys.\ Rev.\ B
{\bf 76}, 165307 (2007).

\bibitem{Chu2008} B. Zhou, H.-Z. Lu, R.-L. Chu, S.-Q. Shan, and Q.
Niu, Phys.\ Rev.\ Lett.\ {\bf 101}, 246807 (2008).

\bibitem{Qiao2010} Z. Qiao, S. A. Yang, W. Feng, W.-K. Tse, J.
Ding, Y. Yao, J. Wang, and Q. Niu, Phys.\ Rev.\ B {\bf 82},
161414(R) (2010).


\bibitem{Yang2011} Y. Yang, Z. Xu, L. Sheng, B. Wang, D. Y. Xing,
and D. N. Sheng, Phys.\ Rev.\ Lett.\ {\bf 107}, 066602 (2011).
\bibitem{Chen2011} T.-W. Chen, Z.-R. Xiao, D.-W. Chiou and G.-Y.
Guo, Phys.\ Rev.\ B {\bf 84}, 165453 (2011).



\bibitem{Ma2015} E. Y. Ma, M. R. Calvo, J. Wang, B. Lian,  M. Muhlbauer,  C. Brune,
Y.-T. Cui,  K. Lai,   W. Kundhikanjana,  Y. Yang, M. Baenninger,
M. K\"{o}nig,  C. Ames,  H. Buhmann, P. Leubner, L. W. Molenkamp,
S.-C. Zhang, D. Goldhaber-Gordon, M. A. Kelly, and Z.-X. Shen,
Nature Comm.\ {\bf 6}, 7252 (2015).


\bibitem{Shao2008} L. B. Shao, S.-L. Zhu, L. Sheng, D. X. Xing,
and Z. D. Wang, Phys.\ Rev.\ Lett.\ {\bf 101}, 246801 (2008).

\bibitem{Liu2013} X.-J. Liu, K. T. Law, T. K. Ng, and Patrick A.
Lee, Phys.\ Rev.\ Lett.\ {\bf 111}, 120402 (2013).

\bibitem{Jo2014} G. Jotzu, M. Messer, R. Desbuquois, M. Lebrat, T. Uehlinger, D. Greif, and T.
Esslinger, Nature, {\bf 515}, 237 (2014).

\bibitem{Zarea2009} M. Zarea and N. Sandle, Phys.\ Rev.\ B {\bf 79},
165442 (2009).





\bibitem{Hsiang2001} W.-Y. Hsiang and D.-H. Lee, Phys.\ Rev.\ A {\bf
64}, 052101 (2001).

\bibitem{Sticlet2012} D. Sticlet, F. Piechon, J.-N. Fuchs, P.
Kalugin, and P. Simon, Phys.\ Rev.\ B {\bf 85}, 165456 (2012).


\bibitem{Murakami2008}S. Murakami, Prog.\ Theor.\ Phys.\ Suppl.\ {\bf 176}, 279
(2008); M. Oshikawa, Phys.\ Rev.\ B {\bf 50}, 17357 (1994);Y.
Hatsugai, M. Kohmoto, and Y.-S. Wu, Phys.\ Rev.\ B {\bf 54}, 4898
(1996).

\bibitem{Yang2015} Bo Yang, Phys.\ Rev.\ B {\bf 91}, 241403(R)
(2015).




\end{thebibliography}
\end{document}